\documentclass[a4paper,fleqn,usenatbib]{mnras}

\usepackage[T1]{fontenc}
\usepackage{ae,aecompl,psfrag}

\def\apj{ApJ}
\def\mnras{MNRAS}

\def\aap{A\&A}                   
\def\apjs{ApJS}                  
\def\pasp{PASP}                  
\def\apjl{ApJ}                   

\def\aj{AJ}
\def\nat{Nature}

\def\gtsim{\mathrel{\hbox{\rlap{\hbox{\lower4pt\hbox{$\sim$}}}\hbox{$>$}}}}

\def\Msun{M$_{\odot}$}
\def\ga{{\rm\thinspace gauss}}

\def\Msun{\hbox{$\rm\thinspace M_{\odot}$}}

\def\simless{\mathbin{\lower 3pt\hbox
	{$\,\rlap{\raise 5pt\hbox{$\char'074$}}\mathchar"7218\,$}}} 
\def\simgreat{\mathbin{\lower 3pt\hbox
	{$\,\rlap{\raise 5pt\hbox{$\char'076$}}\mathchar"7218\,$}}} 

\def\h0{\hbox{{\rm H}$^0$}}
\DeclareMathAlphabet{\vib}{OML}{cmm}{m}{it}

\usepackage{graphicx}	
\usepackage{amsmath}	
\usepackage{amssymb}	

\title[Galaxy evolution in protoclusters]{Galaxy evolution in protoclusters}

\author[Muldrew, Hatch \& Cooke]{
Stuart I. Muldrew$^{1}$,
Nina A. Hatch$^{2}$\thanks{E-mail: nina.hatch@nottingham.ac.uk } and
Elizabeth A. Cooke$^{3}$ 
\\
$^{1}$Department of Physics and Astronomy, University of Leicester, Leicester, LE1 7RH, UK\\
$^{2}$School of Physics and Astronomy, University of Nottingham, Nottingham, NG7 2RD, UK\\
$^{3}$Centre for Extragalactic Astronomy, Durham University, South Road, Durham, DH1 3LE, UK
}

\date{Accepted 2017 September 19. Received 2017 September 5; in original form 2017 June 30.}

\pubyear{2017}

\begin{document}
\label{firstpage}
\pagerange{\pageref{firstpage}--\pageref{lastpage}}
\maketitle

\begin{abstract}
We investigate galaxy evolution in protoclusters using a semi-analytic model applied to the Millennium Simulation, scaled to a \textit{Planck} cosmology.  We show that the model reproduces the observed behaviour of the star formation history (SFH) both in protoclusters and the field.  The rate of star formation peaks $\sim0.7\,{\rm Gyr}$ earlier in protoclusters than in the field and declines more rapidly afterwards. This results in protocluster galaxies forming significantly earlier: 80\% of their stellar mass is already formed by $z=1.4$, but only 45\% of the field stellar mass has formed by this time.  The model predicts that field and protocluster galaxies have similar average specific star-formation rates (sSFR) at $z>3$, and we find evidence of an enhancement of star formation in the dense protoclusters at early times. At $z<3$, protoclusters have lower sSFRs, resulting in the disparity between the SFHs. 
We show that the stellar mass functions of protoclusters are top-heavy compared with the field due to the early formation of massive galaxies, and the disruption and merging of low-mass satellite galaxies in the main haloes. The fundamental cause of the different SFHs and mass functions is that dark matter haloes are biased tracers of the dark matter density field:  the high density of haloes and the top-heavy halo mass function in protoclusters result in the early formation then rapid merging and quenching of galaxies.  
We compare our results with observations from the literature, and highlight which observables provide the most informative tests of galaxy formation.
\end{abstract}

\begin{keywords}
galaxies: clusters: general -- galaxies: evolution --  galaxies: formation
\end{keywords}



\section{Introduction}

In the cosmological paradigm of $\Lambda$CDM, dark matter forms haloes within fluctuations in the primordial matter density field. Baryons cool in the centre of these haloes forming stars, which collect to make galaxies. Gravity causes larger structures to assemble; dark matter haloes merge to form larger haloes. Galaxies within those haloes may merge, or they may become satellites that orbit the most massive galaxy within the halo. Today, the most massive haloes in the Universe are hosts to hundreds to thousands of satellite galaxies, and are known as galaxy clusters.

Collapsing from regions of $10-50$ co-moving Mpc across, clusters are representative regions of the Universe, making them ideal tools for observational cosmology (see \citealt{Allen2011} for a review). However, the galaxies that lie within clusters are not representative of the Universe.  The morphological and colour distributions of cluster galaxies are starkly different from field galaxies \citep[e.g.,][]{Dressler1980,Balogh2004,Bamford2009}, therefore galaxies must have evolved differently in protoclusters.  

Some differences, such as the prevalence of S0 galaxies in denser environments, developed relatively recently \citep[$z\le0.4$:][]{Dressler1997}. But differences in the stellar populations of cluster and field galaxies have been found up to $z\sim2$  \citep[e.g.][]{Steidel2005,Gobat2008,Rettura2010,Hatch2011b}, and so must have been instigated early in the Universe.  Clusters are young structures, having only collapsed relatively recently \citep{Chiang2013,Muldrew2015}. Therefore, to investigate the processes that drive the different evolutionary paths of cluster galaxies we must investigate them during their epoch of formation, when they resided within protoclusters. 
 
The processes that cause these differences between cluster and field galaxies are many and varied.  Dark matter haloes are biased tracers of the underlying dark matter distribution \citep{Abell1958,Sheth2002}, therefore the early formation of dark matter haloes in protoclusters may influence the rate of galaxy formation and their subsequent evolution. The efficiency of internal galaxy feedback processes, such as AGN feedback and winds from supernovae and massive stars, may have a dependency on the large scale environment \citep[e.g.,][]{Henriques2017}. In addition, satellite-specific processes, such as ram pressure stripping \citep{Gunn1972} and  galaxy harassment \citep{Moore1998}, may impact the evolution of protocluster galaxies well before they enter the cluster \citep{McGee2009,Berrier2009}. 

The relative significance of these processes remains difficult to determine observationally. Protoclusters are rare and hard to identify in observations, so a large, well-characterised sample has not yet been obtained. On the other hand, cosmological simulations provide a powerful tool to compare the evolutionary histories of galaxies within different environments. They allow for a comparison between cluster and field galaxies at every epoch, with the ability to directly link progenitors and their descendants without reliance on statistical extrapolation. 

In \citet{Muldrew2015}, we used a semi-analytic model (SAM) applied to the Millennium Simulation \citep{Springel2005} to explore the structure of forming clusters and protoclusters. We showed that protoclusters exist in a range of evolutionary states, independent of their future $z=0$ cluster mass, highlighting the importance of large samples in protocluster studies. In this paper we explore how galaxies form within these assembling structures, focussing on the differences between galaxy formation occurring within protoclusters and the field. We aim to investigate where, when and how the differences between  cluster and field galaxies were instigated. 

The influence and timescale of satellite-specific processes that occur in clusters, e.g., ram pressure stripping and galaxy harassment, have been studied in detail using numerical simulations \citep[for a review see ][]{DeLucia2011}. The environmental history of cluster galaxies and cluster assembly history have also been investigated \citep{Bruggen2008,Berrier2009,McGee2009,DeLucia2012}, as has the chemical enrichment of the intra-cluster medium \citep[e.g.,][]{DeLucia2004}. However, most of these models have been optimised to match the observed properties of clusters in the present day, with few constraints on their properties at higher redshifts. Thus, while the end result matches observations, the routes by which galaxies form and evolve are not well constrained and may be incorrect.

In this work we investigate protocluster galaxy evolution using the \citet{Henriques2015} SAM, which not only reproduces the present day galaxy stellar mass function (GSMF), but also agrees reasonably well with the GSMF up to $z=3$. We compare the predicted protocluster galaxy evolution models to observations of protoclusters in the literature whenever possible to test the reliability of the model. We also highlight where the theory of galaxy formation can be improved, or is not well constrained by the observations.

The galaxy formation model of \citet{Henriques2015} has some notable tensions with current observations: the shape of the cosmic star formation history and the abundance of passive satellite galaxies. For our paper, poorly constrained satellite-specific processes, such as ram pressure stripping, tidal disruption and merger rates, may result in erroneous conclusions regarding the root cause of the divergence of cluster and field galaxy evolution. We therefore use the models to highlight which observables provide the best tests of the galaxy formation model, with the vision that future missions, such as the James Webb Space Telescope ({\it{JWST}}), LSST and {\it{Euclid}}, will be able to robustly observe these systems and reveal the details of cluster galaxy formation.  

The structure of the paper is as follows: Section \ref{sec:method} outlines the SAM we use throughout this study. Section \ref{sec:results} presents our results, including: the star formation history in protoclusters (\S\ref{sec:sfh}); how, when and where protoclusters galaxies are quenched (\S\ref{sec:quenched}); the growth of stellar mass in protoclusters and the efficiency in forming galaxies (\S\ref{sec:SM}); the evolution in the shape of the GSMF (\S\ref{sec:SMF}). In Section \ref{sec:discussion} we examine the cause of the difference between the protocluster and field galaxies and present predictions and tests for future observational studies of high redshift clusters. Our conclusions are in Section \ref{sec:conclusions}.

\section{Methods}
 \label{sec:method}
 
For this study we require a large sample of galaxy clusters whilst also sampling a low enough mass in the GSMF. Therefore we use the SAM of \citet{Henriques2015} applied to the Millennium Simulation \citep{Springel2005}, scaled to a \textit{Planck} cosmology \citep{Planckcosmology2014}.  Following the method of \citet{Angulo2015}, an updated version of \citet{Angulo2010}, the original Millennium Simulation was scaled to correspond to the following cosmological parameters: $\sigma_{\rm 8}=0.829$, $H_{\rm 0}=67.3\,{\rm km\,s^{-1}\,Mpc^{-1}}$, $\Omega_{\rm \Lambda}=0.685$, $\Omega_{\rm b}=0.0487$ and $n=0.96$.  The result of this scaling led to a box of side length $480\,h^{-1}{\rm Mpc}$ comoving.

Friends-of-Friends haloes were identified and processed using \textsc{subfind} \citep{Springel2001}.  Merger trees were then populated with galaxies using the SAM of \citet{Henriques2015}, an updated version of \citet{Guo2011}.  The most pronounced changes to the model were a delay in the reincorporation of wind ejecta, the lowering of the surface density threshold for turning cold gas into stars, limiting ram-pressure stripping to haloes that are more massive than $10^{14}\,h^{-1}{\rm M_{\odot}}$ and a modification to the radio mode AGN feedback.  These changes led to an improved match between the model stellar mass functions and observations and, most importantly for this work, an improved evolution of the stellar mass function between $0<z<3$ (see their figure 2). 

Following on from previous work \citep{Muldrew2015} we implemented a stellar mass cut at all redshifts so that only galaxies with $M_*>10^8\,h^{-1}{\rm M_{\odot}}$ were included in our sample.  We defined galaxy clusters as all haloes with a mass $M_{200} \geq 10^{14}\,h^{-1}{\rm M_{\odot}}$ at $z=0$, yielding a sample of 2136 clusters.  All galaxies within the Friends-of-Friends halo of a cluster were defined as cluster members.  We then defined protocluster galaxies as being any galaxy within the merger tree of a cluster member.  At a given redshift we will refer to the protocluster as the entire ensemble of objects that will merge into the $z=0$ cluster, while we will reserve the term `main halo' for the largest progenitor halo. By definition, all of the protocluster galaxies in our models will have collapsed into the main halo by $z=0$. 

\begin{figure}
 \psfrag{a}[][][1][0]{\large${\rm log}[M_*/(h^{-1}{\rm M_{\odot}})]$}
 \psfrag{b}[][][1][0]{\large${\rm log}(\Phi)$}
 \psfrag{c}[r][][1][0]{\large $z=0$}
 \psfrag{d}[r][][1][0]{\large$z=1$}
\includegraphics[width=90mm]{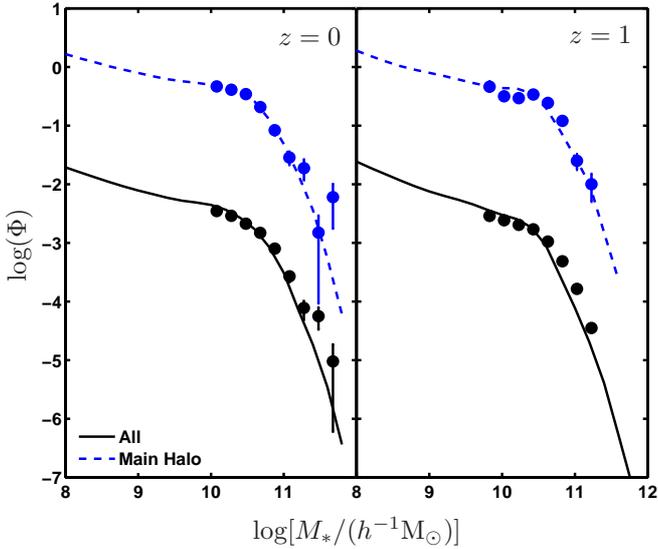}
\caption{\label{fig:MF_data}
The stellar mass functions of field (black solid lines) and main halo (blue dashed lines) galaxies. The main halo mass function has been vertically offset for clarity and the central galaxies have been removed to match the observations. Overlaid are the observed stellar mass functions of field and cluster galaxies (black and blue circles, respectively) from \citet{Calvi2013} at $z=0$ and \citet{vanderBurg2013} at $z=1$. These have been scaled to align with the models to show the agreement in the shapes of the mass functions, and the cluster measurements do not include the brightest cluster galaxies (BCGs). The \citet{Henriques2015} SAM is able to reproduce the shape of the galaxy stellar mass functions at $z=0$ and $z=1$ in both environments. 
}
\end{figure}

The main halo of a protocluster is the structure that can be observed as a high redshift cluster or group and has a compact size of $\sim1$\,Mpc or less. The protocluster consists of hundreds of haloes and extends over tens of comoving Mpc. The main halo is generally only a minor component of the protocluster until $z<0.5$. The fraction of protocluster mass that lies within the main halo ranges between 3 and 20\% at $z=2$, reaching up to 55\% by $z=1$ \citep{Muldrew2015}. 
The SAM dictates that different physical processes affect the evolution of central and satellite galaxies, therefore  we expect galaxy evolution to proceed differently in the main halo compared to whole ensemble of protocluster galaxies due to the different ratio of centrals to satellites in each population. 
At high redshifts ($z>2$) the majority of protocluster galaxies are centrals of their own haloes \citep{Contini2016}. As the protocluster collapses, the centrals gradually become satellites of either the main halo or other large haloes in the protocluster \citep{Contini2016}. On the other hand, all of the galaxies within the main haloes are satellites except for the brightest cluster galaxy (BCG). 

In Section \ref{sec:results} we therefore present results for both the whole protocluster population and the main halo subset. This work aims to help interpret observations of protoclusters, where satellite and central galaxies may not be easily distinguished, therefore we do not remove central galaxies from either the main halo nor protocluster samples, with the exception of Fig.\,\ref{fig:MF_data}.  In order to make a cleaner comparison, we convert all observational parameters where a \citet{Salpeter1955} Initial Mass Function (IMF) is used to a \citet{Chabrier2003} IMF to agree with that used within the simulation.  This is done by dividing the Salpeter parameters (e.g.\,mass) by 1.64.

In Fig.\,\ref{fig:MF_data} we display the GSMF of satellite galaxies in the main haloes to mimic the observations. 
Earlier generations of SAMs displayed a strong dependence of the high mass end of the GSMF on environment at $z<1$ \citep{Vulcani2014}, which was at odds with the observational results of \citet{Calvi2013} and \citet{Vulcani2013}. The \citet{Henriques2015} model solves this tension (Fig.\,\ref{fig:MF_data}) and also provides a good fit to the GSMF from $z=1$ clusters and the field \citep{vanderBurg2013}. 

\begin{figure*}
 \psfrag{a}[][][1][0]{\Large$z$}
 \psfrag{b}[][][1][0]{\Large$z$}
 \psfrag{c}[][][1][0]{\Large$z$}
 \psfrag{d}[][][1][0]{\large${\rm log}[\Sigma{\rm SFR}/({\rm M_{\odot}yr^{-1}})]$}
 \psfrag{e}[][][1][0]{\large${\rm log}[\Sigma{\rm SFR}/\Sigma M_*~/(h\,{\rm yr^{-1}})]$}
  \psfrag{f}[][][1][0]{\large${\rm log}[{\rm SFRD}/(h^{3}{\rm M_{\odot}yr^{-1}Mpc^{-3}})]$}
\includegraphics[width=180mm]{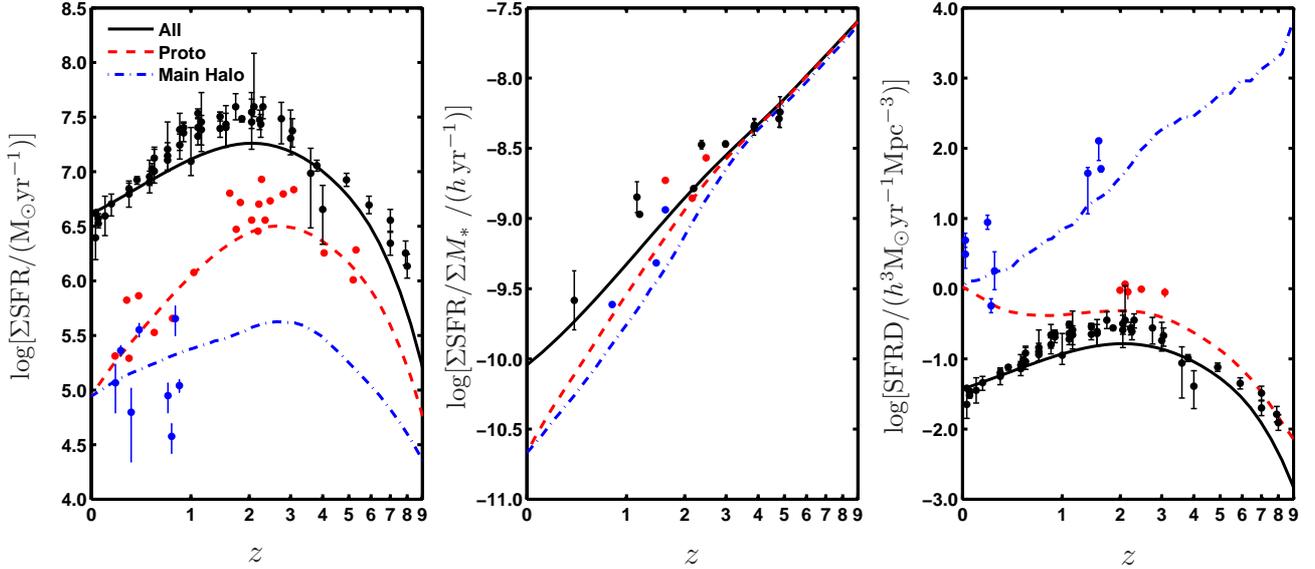}
\caption{\label{fig:CSF}
{\it Left}: (a) The lines show the simulated cosmic star formation history of the field (black solid), protocluster (red dashed) and main halo (blue dot-dashed) galaxies. The curves show the total SFR from these environments within the simulated box. 
Blue circles mark observational SFRs of individual clusters from \citep{Finn2005}, multiplied by a factor of 2136 to correspond to the number of clusters in our simulation. 
The red circles mark observational SFRs of individual protoclusters from \citet{Clements2014, Dannerbauer2014, Casey2016, Hatch2017}. These observations are scaled by a factor of 1200 to match the normalisation of the protocluster SFH in the simulation, see text for a full explanation.  Field data is shown by the black circles, references for which are given in the text. 
{\it Middle}: (b) The lines show the simulated specific SFRs of galaxies within the field (black solid), protoclusters (red dashed) and main haloes (blue dot-dashed). The circles mark observational data of the field (black), protoclusters (red) and clusters (blue) from \citet{Tadaki2011,Tasca2015,Hatch2017}.
{\it Right}: (c) The lines show the SFR density (SFRD) of the field (black solid) and the protocluster (red dashed) in the simulation. The red circles are observational data of protoclusters from \citet{Casey2016}, black circles are field data as in panel (a) and blue circles show cluster data from \citet{Stroe2015,Santos2013,Santos2014,Santos2015}. There is always a higher SFRD in protoclusters and clusters than the field due to the high galaxy density in these regions.
}
\end{figure*}

\section{Results}
\label{sec:results}
\subsection{The star formation history of clusters} 
\label{sec:sfh}

The star formation history (SFH) shows the rate at which baryons are converted into stars as a function of time. Since stars are the factories in which metals are made, and stellar winds and supernovae can heat the surrounding gas haloes, the SFH influences the chemical enrichment and heating of the intergalactic and intracluster medium.  Fig.\,\ref{fig:CSF}a shows the total star formation rate (SFR) as a function of redshift within the simulated box of side length $480\,h^{-1}{\rm Mpc}$. Star formation within the field, protocluster and the main halo environments are shown as black solid, red dashed and blue dot-dashed lines, respectively. 

Clusters assemble 60--75\% of their final mass after $z=1$ \citep{Muldrew2015}. At this time, the SFRs of protocluster galaxies are already an order of magnitude lower than their peak. At all redshifts, the protocluster curve lies far above the main halo curve which means that most of the stars which end up in clusters form within protocluster galaxies outside the main halo.  Therefore, to examine the differing SFHs of cluster and field galaxies we must compare the field curves to those of the protocluster.

The field SFH has been intensively studied with UV and IR surveys, and we overplot the observed SFHs of both the field and clusters in Fig.\,\ref{fig:CSF}a.  The field data comes from  \citet{Sanders2003,Takeuchi2003,Wyder2005,Schiminovich2005,Dahlen2007,Reddy2009,Robotham2011,Magnelli2011,Cucciati2012,Bouwens2012a,Bouwens2012b,Schenker2013,Gruppioni2013}, and is compiled by \citet{Madau2014}. No survey has measured the total SFR within a volume-limited sample of protoclusters, so instead we use measurements of individual protoclusters \citep{Clements2014, Dannerbauer2014, Casey2016, Hatch2017}, and scale by a factor of 1200 to match the normalisation of the simulated protocluster SFH at low redshift. Since there are 2136 protoclusters in the simulation, this suggests that the observed protoclusters have more star formation than the simulated protocluster sample, and therefore may be more massive. The mass distribution of the simulated protoclusters is greatly skewed towards the low mass end ($10^{14}\,h^{-1}$\Msun), whereas observations are more likely to select the most massive protoclusters due to selection effects. Therefore Fig.\,\ref{fig:CSF}a allows us to examine the difference in the shapes of the SFHs, but not the normalisations.  In general there is reasonable agreement between the data and the models for both field and protocluster/cluster regions. 

At early times ($z>4$) the gradients of the SFHs are similar in both environments meaning the rate of increase of star formation is similar.  But after $z=4$, the deceleration of star formation is more rapid in protoclusters, and the SFR peaks at $z=2.64$, compared to the field which peaks at a later time of $z=2.07$. The epoch of peak star formation occurred 0.7\,Gyr earlier within protoclusters than in the field.  

After the earlier peak in the SFR, there is a more rapid decrease. The SFR decreases by a factor of 36 between the peak of SFR at $z=2.64$ and today in protoclusters, whereas the decrease in the field is a more modest factor of 4.  The SFR distribution for field galaxies is broader than for the protocluster galaxies. The full width at half maximum of the SFH is $6.1$ billion years in the field, whereas this period only lasts $3.4$ billion years within protoclusters. This means the stellar population that ends up in clusters also formed over a shorter period of time than the field population.

To investigate why the star formation peaks earlier and extends over a shorter period of time in protoclusters compared to the field, we examine how the protocluster environment affects the rate at which stars are formed. The specific SFR (sSFR) of the field, protocluster and main halo is shown in Fig.\,\ref{fig:CSF}b. This was calculated as the total SFR divided by the total stellar mass within $M_*>10^{8}\,h^{-1}$\Msun\ galaxies in each environment. 

\begin{figure}
 \psfrag{a}[][][1][0]{\large${\rm log}[M_*/(h^{-1}{\rm M_{\odot}})]$}
 \psfrag{b}[][][1][0]{\large Quenching Efficiency}
\includegraphics[width=1\columnwidth]{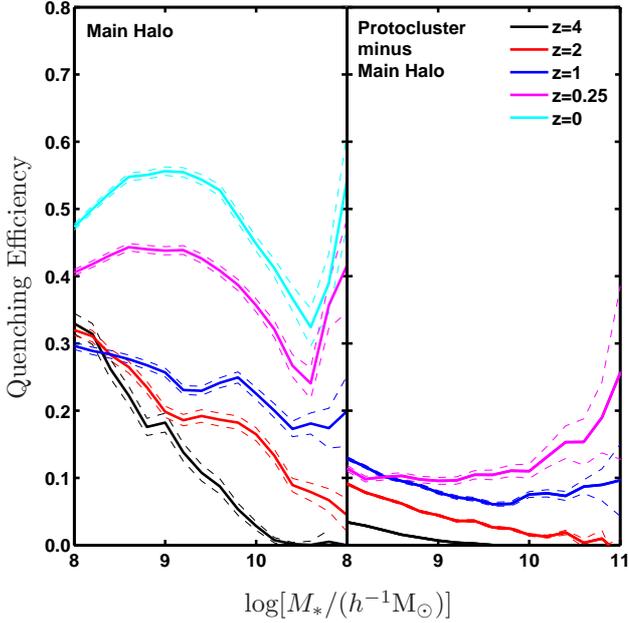}
\caption{\label{fig:qe_cluster}
Quenching efficiency as a function of redshift for both main haloes (left) and the rest of the protocluster (not including the main haloes) (right). The quenching efficiency of the main halo is high, even in the early Universe, however some quenching occurs outside the main halo at all epochs. At $z=0$ there are no protocluster galaxies outside the main halo.  Thin dashed lines correspond to the Poisson error.}
\end{figure}

At $z>3$, stars form at almost the same mass-specific rate in the field, protocluster and main halo. The dense environment has no significant effect on the average sSFR of galaxies at these high redshifts. At $z<3$, the protocluster galaxies, especially those in the main haloes, form stars at a lower rate per unit stellar mass than in the field. This conforms to the usual star formation-density relation that is observed within clusters: galaxies within clusters typically have lower rates of star formation than field galaxies \citep{Kauffmann2004}. This implies that the different SFH is at least partially driven by the suppression of star formation in dense environments. The reason the SFR peaks at an earlier time in protoclusters is because star formation is more suppressed in this environment at $z<3$, rather than being enhanced at earlier times.

We note that these results depend on the physics within the simulation. The parameters governing these processes are often poorly constrained, poorly captured and some processes may be missing entirely. However, it is reassuring that the shape of the simulated protocluster SFH matches the observed shape of the SFH, which suggests that the relative amounts of star formation across cosmic time are approximately correct.

Some recent observations have suggested that star formation may be enhanced within protocluster galaxies (e.g. \citealt{Tran2010}). 
The models never show an enhancement in the SFR per unit mass within dense regions, so the simulation predicts that there is no reversal in the SFR-density relation at high redshift. Observational data points from \citet{Tadaki2011,Tasca2015} and \citet{Hatch2017} are overlaid on the models in Fig.\,\ref{fig:CSF}b. The observations generally show that the sSFR in the field is higher than in clusters and protoclusters at $z<2$, which is in agreement with the simulated trend. However, there are too few data points to conclude whether or not the sSFR is enhanced in dense environments at $z>2$. 

In Fig.\,\ref{fig:CSF}c we show the star formation rate density (SFRD) in the main haloes, protoclusters and the field. We define the volume of the protoclusters and the main haloes as the volume which contains 90\% of the stellar mass in each environment respectively. We find that there is always more SFR per unit volume in protoclusters than the field due to the high mass density within a small volume. Similarly, there is a higher density of star formation in the main haloes than in the entire protocluster regions because the volume of the main halo is much smaller than the protocluster and this region contains the highest stellar mass density. 

We overlay the observational field data shown in Fig.\,\ref{fig:CSF}a, adding cluster/main halo data from \citet{Stroe2015,Santos2013,Santos2014,Santos2015} and protocluster data from \citet{Casey2016}. Again, the data qualitatively agree with the models: protocluster and cluster regions contain a higher density of star formation than the field, but the observations of high-redshift clusters and protoclusters suggest higher SFRDs than predicted by the simulation. 

The simulation shows that protoclusters, especially their main haloes, are cradles of star formation because they contain a high density of galaxies rather than a reversal in the SFR-density relation. In fact, the greatest discrepancy between the field and protocluster SFRD occurs at $z<0.5$, when the protocluster rapidly collapses into a compact cluster-sized halo \citep{Muldrew2015}, which causes the red curve in this plot to increase after this epoch. The SFRD within the main halo constantly decreases with redshift because the volume of the main haloes is constantly increasing with redshift. Although Fig.\,\ref{fig:CSF}a shows that the total star formation in the main haloes increases from $z=9$ to $z=2.7$, the dominant effect on the SFRD is the growth of the halo volume, causing the SFRD to decrease with time.

In summary, the SFH of cluster and field galaxies differs: the SFR  peaks earlier and extends over a shorter period of time in protoclusters than in the field. This is due to enhanced levels of quenching of star formation since at least $z=3$.  The simulation suggests there is no reversal in the SFR-density relation at high redshift. Due to cluster-to-cluster variation in observations there may be indications of such a reversal in individual clusters; however, the simulation suggests that stars form at a similar mass-specific rate in all environments at $z>3$ on average, and at lower rates in $z<3$ clusters and protoclusters. 

\subsection{Quenching of cluster galaxies}
\label{sec:quenched}
In Section \ref{sec:sfh} we showed that the formation of stars is suppressed in protoclusters compared to the field from at least $z\sim3$, and this suppression becomes more significant with time. Here we investigate where this occurs within protoclusters, which galaxies are affected, and what processes cause the suppression of star formation. There are few observations that allow us to constrain the timescale and efficiency of the processes that simulate the suppression of star formation, so in this Section we focus on identifying observables that may help constrain these processes.

\subsubsection{Where is star formation suppressed in protoclusters?}

\begin{figure*}
 \psfrag{a}[][][1][0]{\Large${\rm log}[M_*/(h^{-1}{\rm M_{\odot}})]$}
 \psfrag{b}[][][1][0]{\Large Quenched fraction}
 \psfrag{c}[l][][1][0]{\Large$z=3.5$}
 \psfrag{d}[l][][1][0]{\Large$z=2.25$}
 \psfrag{e}[l][][1][0]{\Large$z=1.25$}
 \psfrag{f}[l][][1][0]{\Large$z=0.35$}
\includegraphics[width=180mm]{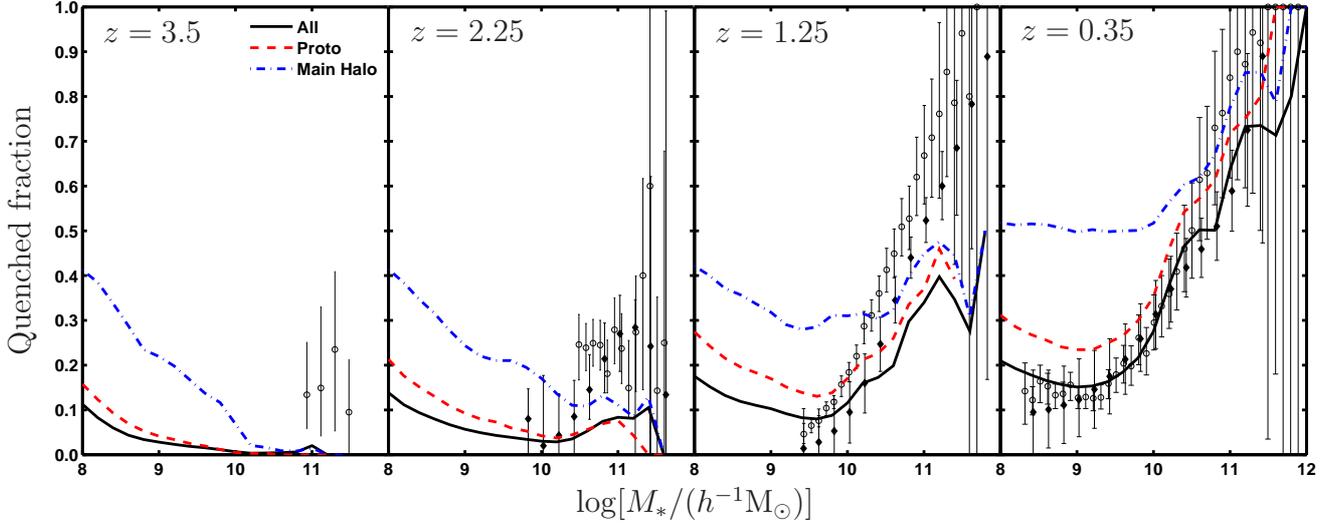}
\caption{\label{fig:Passfrac}
The fraction of galaxies in the field (black solid), protocluster (red dashed) and main halo (blue dot-dashed) that are classified as passive based on their low sSFR. Open circles and filled diamond data points show the observed quenched fraction in the field, defined from rest-frame $UVJ$ colours \citep{Muzzin2013b} and $NUV,R,J$ colours \citep{Ilbert2013}, respectively.
The model predicts that at high redshift tidal stripping of satellite galaxies is the main process of star formation quenching in dense environments. At lower redshifts AGN feedback processes in intermediate and high mass galaxies play a greater role in quenching. 
}
\end{figure*}

We classify galaxies as star-forming or passive based on their sSFR. At $z=0$ we set the sSFR boundary between the two subsets as ${\rm log}[{\rm sSFR}/(h\,{\rm yr^{-1}})]=-11$ \citep[as commonly used in the literature, e.g.][]{Wetzel2012} and measure the difference between the peak in the field sSFR distribution and this value. At higher redshifts the sSFR peak is at higher values so we define the star-forming/passive boundary by keeping the separation to the peak sSFR fixed, yielding values of ${\rm log}[{\rm sSFR}/(h{\rm yr^{-1}})]=\{-10.78,-10.35,-10.01, -9.69\}$ at $z=\{0.35, 1.25, 2.25, 3.5\}$. 

To examine whether galaxies are quenched uniformly throughout the protocluster we display the quenching efficiency of galaxies within the main haloes and within the outer protocluster structures (excluding the galaxies within the main haloes) in Fig.\,\ref{fig:qe_cluster}.
 We define the quenching efficiency as:
\begin{equation}\label{eqn:QE}
QE=\frac{f_q-f_{q,all}}{1-f_{q,all}}
\end{equation}
where $f_q$ is the fraction of quenched galaxies in the dense environment and $f_{q,all}$ is the fraction of quenched galaxies in the field (i.e. all galaxies in the simulation box).  This quenching efficiency shows the influence of environmental processes on quenching of star formation within the galaxy population. 

Fig.\,\ref{fig:qe_cluster} reveals that quenching within the main halo is more efficient than in the rest of the protocluster. The left panel of Fig.\,\ref{fig:qe_cluster} shows that the efficiency of the quenching processes is negatively correlated with stellar mass at M$<10^{10.5}$\Msun\ so low mass galaxies are more likely to be quenched than high mass galaxies. 

At $z<1$ there is an increase in the QE of high mass galaxies (M$>10^{10.5}$\Msun). This is due to an increase in quenched massive galaxies in the field at these redshifts (see right-hand panel of Fig.\,\ref{fig:Passfrac}) which is caused by efficient AGN-driven quenching. The $f_q-f_{q,all}$ term in equation \ref{eqn:QE} remains approximately constant at M$>10^{10.5}$\Msun, but the $1-f_{q,all}$ term in the denominator rapidly decreases with mass, resulting in a rapid increase in the QE of massive galaxies at $z<1$.  
The increase in the importance of AGN feedback in the field relative to the protocluster/cluster is confirmed by observations that show the relative density of AGN in the field increases rapidly compared to the AGN density in clusters after $z=1$ \citep{Martini2013}. 
 
The right-hand panel of Fig.\,\ref{fig:qe_cluster} reveals that not all quenching occurs in the main halo. The QE in the protocluster regions outside the main haloes is a few to $\sim30$\%, but is lower than the main halo QE at each redshift. The main reason for the different QEs is the different central to satellite galaxy ratio of the two environments. As discussed in the following section, satellite galaxies in the model are efficiently quenched by processes that do not affect central galaxies.  The main halo population is dominated by satellite galaxies as each halo only hosts one central BCG.  Conversely, protoclusters consist of hundreds of haloes and most protocluster material does not reside within massive groups. Instead many protocluster galaxies are centrals of their own halo \citep{Contini2016,Hatch2016}, so the average QE of the protocluster galaxies beyond the main halo is relatively low. Galaxy ``pre-processing", where galaxies are affected by their environment before they enter the main halo, may affect a significant fraction of the {\it surviving} satellite galaxies in clusters and we refer the reader to \citet{Berrier2009,McGee2009} and \citet{DeLucia2012} for a full discussion of the importance of galaxy pre-processing in clusters. 

\subsubsection{Why do galaxies quench in the main haloes?}

\begin{figure*}
 \psfrag{a}[][][1][0]{\Large$z$}
 \psfrag{b}[][][1][0]{\large${\rm log}[M_*/(h^{-1}{\rm M_{\odot}})]$}
  \psfrag{c}[][][1][0]{\Large$z$}
 \psfrag{d}[][][1][0]{\large$M_*/{ M_{*, z=0}}$}
  \psfrag{e}[][][1][0]{\Large$z$}
 \psfrag{f}[][][1][0]{\large$M_*/{M_{DM, z=0}}$}
\includegraphics[width=180mm]{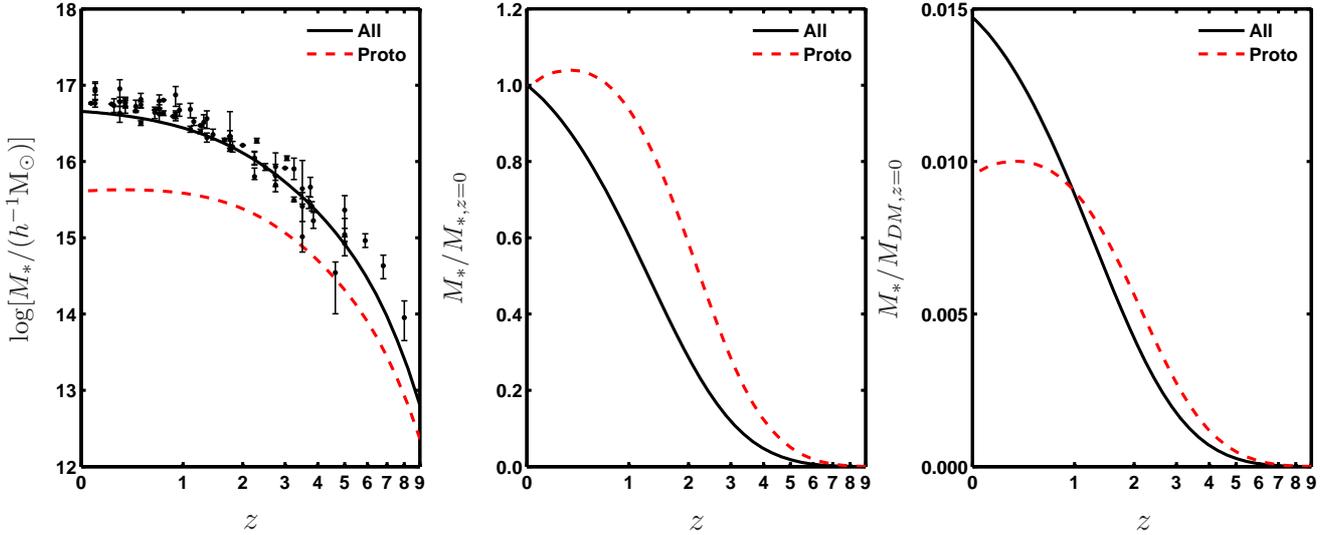}
 \caption{\label{fig:sm_total}Left: (a) Total stellar mass in the field (black solid line) and protocluster (red dashed line) as a function of redshift. Data points are observations in the field, see text for references. The simulation was optimised to reproduce the stellar mass function at $0<z<3$ and thus matches the data well.
Middle: (b) Total stellar mass normalised by the total galactic stellar mass in each environment at $z=0$. Since the total heavy element mass is given by $Z(z)=yM_*(z)$, this evolution also shows the ratio of the total mass in metals to the final mass in galactic metals for each environment at $z=0$. 
Right: (c) Total stellar mass normalised by the total dark matter mass in each environment at $z=0$.}
\end{figure*} 
The rate of star formation in the simulation is determined by the amount of cold gas within a galaxy. Galaxies cease forming stars when they do not have sufficiently large cold gas reservoirs. The cold gas can be removed from a galaxy by supernovae and stellar wind feedback, or ram pressure stripping. The cold gas can also be depleted by star formation and no further rejuvenation occurs if the hot halo has been tidally stripped, which halts further infall of gas onto the galaxy, or if radio-mode AGN feedback prevents the hot gas from cooling onto the cold disc. 

Fig.\,\ref{fig:Passfrac} shows the fraction of quenched galaxies as a function of stellar mass for four redshift snapshots in the SAM together with observational data from \citet{Muzzin2013b} and \citet{Ilbert2013}, who distinguish between star forming and passive galaxies using rest frame $UVJ$ and $NUV,R,J$ colours, respectively. The discrepancies between the models and the observations is primarily due to the different methods used to classify passive galaxies\footnote{The \citet{Henriques2015} model uses the \citet{Muzzin2013b} and \citet{Ilbert2013} passive fractions as constraints for their SAM, and therefore the SAM results are well matched to the observations (see figure 5 in \citealt{Henriques2015}). However, the simulated  galaxies have incorrect colours because of shortcomings in the population synthesis/dust modelling, so \citet{Henriques2015} modify the observational $UVJ$ and  $NUV,R,J$ criteria to separate star forming and passive galaxies in their simulation to match the observations.}, therefore the observations should only be compared qualitatively with the models. At all redshifts the quenched fraction of protocluster galaxies is higher than the field, consistent with observations  \citep{Wetzel2012,Cooke2016,Nantais2017}. 

We find that the quenched galaxies at $z=3.5$ are almost all satellites whose dark matter and hot gaseous haloes have been stripped by tidal interactions. Without a hot halo, the simulation restricts any recycling of gas and infall of primordial gas. Therefore these passive galaxies must have either used up their cold gas supply, or stellar feedback has removed the remaining cold gas so the galaxies no longer 
form new stars. The negative correlation between the passive fraction and the stellar mass of satellite galaxies suggests that the efficiency of this mechanism is correlated with the mass of the satellite galaxy.

By $z=2.25$, these tidally-stripped galaxies with no rejuvenation are still the primary class of quenched galaxy at low {($\lesssim10^{9}\,h^{-1}{\rm M_{\odot}}$)} masses, but there are also populations of intermediate ($\sim10^{10}\,h^{-1}{\rm M_{\odot}}$) and high {($>10^{10.5}\,h^{-1}{\rm M_{\odot}}$)} mass galaxies which have retained their hot gas haloes and yet are quenched.  AGN-feedback is the only method in the simulation that can limit the cold gas reservoir in these galaxies and cease star formation.  Ram pressure stripping is not operating efficiently in the model until $z\le1$ as the simulation only allows ram pressure stripping to remove hot haloes in host haloes of $M>10^{14}h^{-1}$\Msun. At  $z=1$ only 10\% of the dark matter mass lies within such high mass haloes, so ram pressure stripping is not the main cause of star formation quenching at high redshift in the simulation.

At $z<1.25$, the growth in the passive fraction of intermediate mass galaxies is indicative of additional physical processes quenching galaxies. These passive satellites have retained their subhaloes, unlike the low mass passive satellites at $z>1.25$.   The rise in the abundance of quenched galaxies around $\sim10^{10}\,h^{-1}{\rm M_{\odot}}$ is triggered by AGN feedback.  As galaxies, and hence their black holes, grow,  AGN feedback plays a larger role in quenching satellite galaxies.  
From $z=1.25$ to $z=0$ main haloes also start to obtain masses in excess of $10^{14}\,h^{-1}{\rm M_{\odot}}$ and so ram-pressure stripping begins to play an increasingly important role in the simulation.

The model predicts that the predominant mode of quenching in clusters has evolved with time. At $z>1$, low mass galaxies are quenched when they become satellites. Their hot haloes are tidally stripped and their cold gas reservoirs are either used up through forming stars, or removed by supernovae/stellar wind feedback. Without a hot halo they are unable to rejuvenate their gas supply and they become passive.  High mass galaxies at this epoch are quenched by AGN. At $z<1$, galaxies are still quenched through AGN feedback and tidal stripping, but also increasingly by ram pressure stripping and gas exhaustion. Observationally determining the quenched fraction of protocluster and cluster galaxies at $z>1$, in particular at low galaxy masses ($M_*<10^{9}\,h^{-1}$\Msun), can provide a stringent constraint on the processes that suppress star formation.

\subsection{Growth of stellar mass}
\label{sec:SM}

\begin{figure}
 \psfrag{a}[][][1][0]{\Large${\rm log}[M_*/(h^{-1}{\rm M_{\odot}})]$}
 \psfrag{b}[][][1][0]{\Large${\rm log}(N)$}
\includegraphics[width=1\columnwidth]{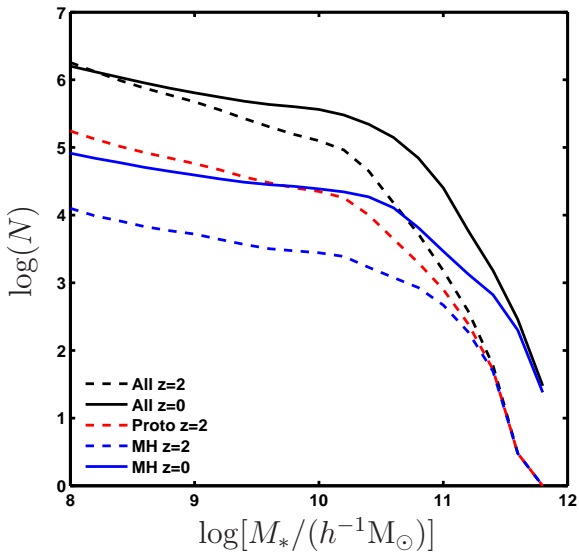}
\caption{\label{fig:smf_num}
The evolution of the galaxy stellar mass functions in the field (black), protocluster (red) and main halo (blue).  Dashed lines correspond to $z=2$ and solid lines to $z=0$. There is significant evolution in the field at intermediate and high masses between $z=2$ and $z=0$, causing a flattening of the low-mass slope. The protocluster galaxies at $z=2$ (red dashed line) evolve into the main halo galaxies at $z=0$ (blue solid line), meaning low mass galaxies are destroyed in the protocluster at $2>z>0$ via merging and tidal disruption.}
\end{figure}

In Section \ref{sec:sfh} we examined the formation of stars in different environments. We showed that protocluster SFRs peaked earlier in time than field galaxies. In this Section we study the build up of stellar mass in the two environments to determine when the bulk of their galactic stellar mass was formed. 

Fig.\,\ref{fig:sm_total}a shows the total stellar mass within galaxies, $M_{*}(z)$, within the simulation box. The circles show the observed stellar mass density of the field from \citealt{Li2009,Gallazzi2008,Moustakas2013,Bielby2012,Perez-Gonzalez2008,Ilbert2013,Muzzin2013b,Arnouts2007,Pozzetti2010,Kajisawa2009,Marchesini2009,Reddy2012,Caputi2011,Gonzalez2011,Lee2012,Yabe2009} and \citealt{Labbe2013}, compiled by \citet{Madau2014}, and scaled by the simulated box volume. The models match the data well because the simulation was optimised to match the stellar mass function from $z=3$ to the present day. 

At very early times, $z>15$, the simulation shows that all of the stellar mass within $M_{*}>10^{8}h^{-1}$\Msun\ galaxies resides in protoclusters. This is due to the early collapse of dark matter haloes in the densest regions of the Universe (i.e., protoclusters), and the subsequent formation of stars within those haloes. We refer the reader to  \citet{Chiang2017} for a discussion of the influence this has on the epoch of reionisation. This period does not last long ($\sim120$\,Myr) and by $z=11.5$ there is a greater amount of stellar mass outside protoclusters than within protoclusters. 

Fig.\,\ref{fig:sm_total}b displays the fraction of final stellar mass that has formed by a given redshift. It reveals that the bulk of stars in cluster galaxies formed early: field galaxies formed the bulk (75\%) of their stellar mass at $z<2$, whereas the same amount of mass is formed at $z>1.6$ in clusters. This is consistent with numerous observations that place the average formation redshift of massive cluster galaxies to be $<z_{\rm form}>=2-3$ \citep[e.g.,][]{Bower1992,Holden2005,Cooke2015}, as well as the difference in the mean stellar ages of cluster and field ellipticals  \citep{Gobat2008,Rettura2010}. 

The growth of stellar mass within protocluster galaxies halts and then reverses at $z=0.35$ because the rate of galaxies being tidally disrupted in the protocluster is similar to the SFR 
at $z<1$, and becomes larger than the SFR at $z=0.35$. Tidal disruption of the galaxies converts galactic stellar content into the intra-cluster component that is observed as intra-cluster light. Negligible amounts of intra-cluster stars are present at $z=1$, increasing to approximately 20-30\% of the total stellar mass within clusters at $z=0$ \citep{Contini2014}. The effect of this stripping on the GSMF is explored in Section \ref{sec:SMF}.

 The total sum of heavy metals produced in each environment is $Z=yM_{*}$, where $y=0.045^{+0.036}_{-0.028}$ is the \lq metal yield\rq\ factor \citep{Henriques2015}.  Therefore the metal production per unit stellar mass will follow the approximate evolution in Fig.\,\ref{fig:sm_total}b scaled by the constant factor of $y$, although the late-time evolution ($z\lesssim1$) of the mass growth does not continue to trace the metallicity evolution since the increasing dominance of stripping causes stellar mass loss from galaxies, even though low-level star formation continues in protoclusters.

\begin{figure*}
 \psfrag{a}[][][1][0]{\Large${\rm log}[M_*/(h^{-1}{\rm M_{\odot}})]$}
 \psfrag{b}[][][1][0]{\Large${\rm log}[N/M_{\rm tot}~/(h^{-1}{\rm M_{\odot}})]$}
 \psfrag{c}[][][1][0]{\Large${\rm log}[M_*/(h^{-1}{\rm M_{\odot}})]$}
 \psfrag{d}[][][1][0]{\Large${\rm log}[N]$}
\includegraphics[width=180mm]{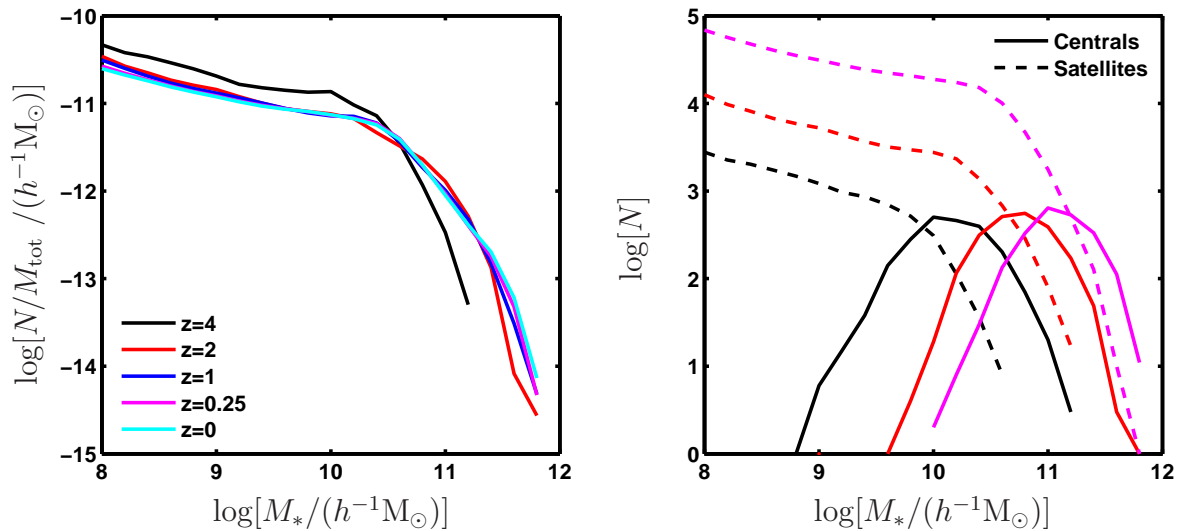}
\caption{\label{fig:SMFtype}
Evolution of the galaxy stellar mass functions for the main haloes.  The left panel shows the stellar mass function for all main halo galaxies, normalised by total stellar mass. The right panel shows the raw number of galaxies in main haloes separated into centrals and satellites at $z=4$, 2 and 0.25.
}
\end{figure*}

Fig.\,\ref{fig:sm_total}b suggests that the metals within protoclusters were formed much earlier than those in the field.  At $z=2$ protoclusters were still tens of co-moving Mpc across and collapse had not yet begun in earnest, yet more than half of cluster metals had formed. This early metal production means that protoclusters were able to pollute the proto-intra-cluster medium (proto-ICM) relatively evenly and these metals would be efficiently mixed as the proto-ICM collapsed to form the dense ICM. These predictions of the metal enrichment of the ICM are consistent with the observational measurements of uniform metallicity in the outskirts of the ICM \citep[e.g.][]{Werner2013,Simionescu2015, Simionescu2017} as well as the lack of evolution in the global ICM metallicity since $z=1.5$ \citep{McDonald2016,Mantz2017}, which suggests that at least 60\% of the metals within clusters were formed by $z=1$.

Fig.\,\ref{fig:sm_total}c shows the total stellar mass within galaxies scaled by the total dark matter halo mass at $z=0$.  At high redshift, $z>1$, protoclusters are more efficient at producing stars compared to the field, as shown by the higher values of $M_*/{M_{DM, z=0}}$. This is due to the earlier formation of dark matter haloes in the densest regions, and the subsequent formation of galaxies within those haloes (see Section \ref{sec:DM}). After $z=1$, the total amount of stars in protocluster galaxies does not greatly increase. This is partly due to the low SFR in protoclusters (see Fig.\,\ref{fig:CSF}) and partly due to the tidal disruption of protocluster galaxies which transfers stars into the intra-cluster halo. The formation of the intra-cluster light is model-dependent and we refer the reader to \citet{Contini2014} for a detailed discussion. 
 
At $z<1$, stars continue to form at a higher sSFR in the field than in protoclusters, so by $z=0$ a higher fraction of baryons in the field have been converted into stars and remain in galaxies. At $z=0$, cluster galaxies have $\sim$33\% less stars than there are in field galaxies that trace the same total dark matter halo mass. Therefore, cluster galaxies are overall less efficient at converting their baryons into galactic stellar matter and retaining it, yet they take a short time to do so. 

\subsection{Stellar mass functions}
\label{sec:SMF}

We examine the distribution of galactic stellar mass in Fig.\,\ref{fig:smf_num} where we display the GSMF at $z=2$ and $z=0$ in the field, protoclusters and the main haloes.  The  normalisation shows the total number of galaxies within each environment in the entire simulated box, of side 480\,$h^{-1}$Mpc. The protocluster and main halo GSMFs are always top-heavy compared to the field. This agrees with observations of $z\sim1$ clusters  \citep{vanderBurg2013}, and protoclusters at $z\sim2.5$ that contain higher fractions of massive galaxies than the field  \citep{Steidel2005, Hatch2011b, Cooke2014}.

Fig.\,\ref{fig:smf_num} shows that the evolution of the GSMF of protocluster galaxies differs from field galaxies. The field galaxies at $z=2$ (black dashed line) are the progenitor population for the $z=0$ field population (black solid line). The number of $10^8\,h^{-1}$\Msun\ galaxies in the field is approximately the same at $z=2$ as $z=0$, but the low-mass slope of the mass function flattens with time due to an increase in the number of massive galaxies. Stars form more efficiently in $\sim10^{12}\,h^{-1}$\Msun\ haloes than in lower mass haloes \citep{Behroozi2013}, building up the mass function at intermediate masses and flattening the low-mass field slope with time.

The progenitor population of the $z=0$ main halo/cluster galaxies (blue solid line) is the protocluster population at $z=2$ (red dashed line). By definition, all the galaxies in the $z=2$ protoclusters end up in the $z=0$ clusters.  Contrary to the field, there are significantly more low-mass galaxies in the $z=2$ protoclusters than in the descendant main haloes at $z=0$. These low mass galaxies are destroyed in the main halo by merging with other galaxies as well as being shredded and becoming part of the intra-cluster light. The fraction of galaxies that merge with the BCG  relative to the fraction that are tidally disrupted depends on the model parameters and current observations of the fraction of intra-cluster light and merger rates in clusters are not sufficient to constrain these parameters. The model suggests that tidal disruption is insignificant between $z=2$ and $1$ compared to the rate of merging galaxies, but tidal disruption is much more significant at $z<1$.

Fig.\,\ref{fig:smf_num} shows that the distribution of galactic mass developed differently in clusters than in the field, and it is merely a coincidence that the field and cluster satellite mass functions are similar today \citep{Calvi2013,Vulcani2013}.  The simulation suggests that protocluster GSMFs are top-heavy compared to the field due to the early formation of large dark matter haloes and their central galaxies. The stellar matter within protoclusters forms comparatively early in a large number of small galaxies, and is gradually redistributed to larger galaxies and the intra-cluster light.  In an average $z=2$ protocluster there are approximately 380 galaxies with $M_*>10^{8}h^{-1}$\Msun, which decreases to only 260 galaxies by $z=0$. This redistribution of stellar mass does not occur to such a great extent in the field, resulting in the different evolution of the protocluster and field GSMFs.

\begin{figure*}
 \psfrag{a}[][][1][0]{\large${\rm log}[M_{200}/(h^{-1}{\rm M_{\odot}})]$}
 \psfrag{b}[][][1][0]{\large${\rm log}[\Phi/(h^{3}{\rm Mpc^{-3}})]$}
 \psfrag{c}[r][][1][0]{\large$z=4$}
  \psfrag{d}[r][][1][0]{\large$z=2$}
   \psfrag{e}[r][][1][0]{\large$z=1$}
    \psfrag{f}[r][][1][0]{\large$z=0.25$}
\includegraphics[width=2\columnwidth]{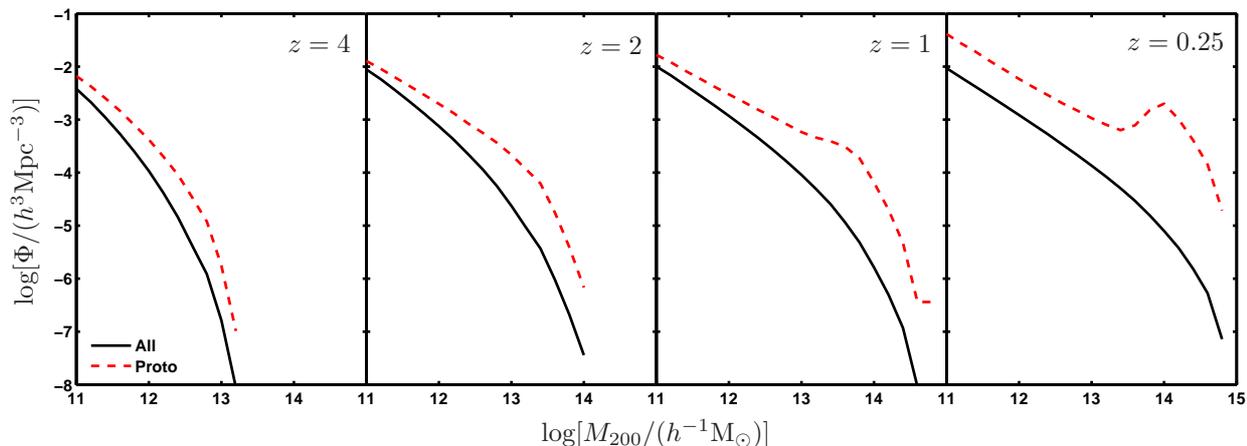}
\caption{\label{fig:HMF}
The dark matter halo mass functions of field (black solid) and protocluster (red dashed) regions. The volume of the protoclusters is taken as the region encompassing 90\% of the protocluster mass. There is a higher density of dark matter haloes in protoclusters at all redshifts and the distribution is top-heavy. The mass gap between the main and other protocluster haloes becomes more apparent at $z\le1$, resulting in the bump at $M_{200}\sim10^{14}\,h^{-1}$\Msun.}
\end{figure*}
The normalisation of the main halo GSMF increases from $z=2$ to $z=0$, primarily due to the accretion of galaxies from the surrounding protocluster into the main halo, but there is no change in the shape of the GSMF. We highlight this further in Fig.\,\ref{fig:SMFtype}, showing the evolution from $z=4$ to $z=0$ for the main halo population. \citet{Nantais2016} showed that the shape of the GSMF of $z=1.6$ clusters is similar to that of $z=1$ clusters suggesting that the evolution of the GSMFs in the simulation is correct\footnote{However, \citet{Nantais2017}  removed the BCGs from the $z=1$ clusters but not the $z=1.6$ clusters so this comparison should be treated with caution.}. 

The right panel of Fig.\,\ref{fig:SMFtype} shows that the shape of the GSMF does not evolve from $z=2$ to $z=0$ due to two competing processes. Over time, protocluster galaxies accrete onto the main halo. This increases the abundance of the satellite population (whilst the central population remains the same size) which is seen as the increase in the normalisation of the satellite GSMF relative to the central GSMF. In addition, more massive galaxies become satellites of the clusters with time, which is seen as the increase in mass of the knee of the satellite GSMF and the decrease in the mass difference between the most massive satellites and the centrals.  At the same time, the satellites merge with the central galaxies resulting in the growth of the central BCGs. Thus as the main halo grows, the central galaxy becomes more massive, but also becomes less dominant (in terms of number) in the main haloes. 

The simulated GSMF of the main haloes does not change shape because the accretion of satellite galaxies and the growth of the central galaxies are related \citep[see supplementary material regarding the merger rates in ][]{Henriques2015}. We note that at $z>2$ the GSMF of the main haloes does evolve, with the knee of the mass function becoming progressively more massive with time. This is because the central galaxies of the main haloes grow rapidly due to star formation, which is not correlated to the accretion rate of satellite galaxies. 
Thus the evolution of the main halo GSMF provides insight into the relative growth of the central galaxies compared to the accretion rate of the haloes.

\section{Discussion}
\label{sec:discussion}

\subsection{The effect of the dense dark matter density field in protoclusters}
\label{sec:DM}

The simulation suggests that cluster galaxies differ from field galaxies because they have different star formation and assembly histories. Cluster galaxies experience more mergers and tidal disruption, and earlier suppression of star formation. The physics governing these processes is the same throughout the simulation, so the major physical difference between the cluster and field regions is the underlying dark matter halo distribution, and how galaxies populate those haloes. To explore the different dark matter distributions we compare the evolution of the protocluster and field halo mass function in Fig.\,\ref{fig:HMF}. 

Fig.\,\ref{fig:HMF}  shows that dark matter haloes are more abundant and top-heavy in protoclusters compared to the field at all redshifts. This is because protoclusters occupy the most extreme perturbations in the primordial density field, and dark matter haloes are biased tracers of the dark matter density field \citep{Abell1958,Tinker2010}. The high abundance of dark matter haloes promotes the growth of protocluster galaxies for two reasons. First, galaxies form efficiently because the baryonic density is only high enough for star formation in the cores of dark matter haloes. Second, the compact configuration of haloes results in an efficient merging halo rate and subsequently, more rapid galaxy assembly.

The high abundance and top-heavy halo mass functions also act to suppress galaxy growth. The top-heavy mass functions mean that minor mergers are more common in protoclusters than in the field. The time for a satellite galaxy to merge with the central galaxy of a halo is proportional to the ratio of host to satellite halo mass, so minor mergers spend a long time as satellite galaxies and are affected by satellite-specific quenching processes, such as ram pressure stripping. Quenching due to tidal stripping (as described in \S\ref{sec:quenched}) is also more efficient in protoclusters because the tidal radius is proportional to the ratio of satellite to host halo mass. Hence the abundance of minor mergers in protoclusters results in efficient tidal stripping, leading to quenching of star formation and tidal disruption.

The high abundance of massive haloes in protoclusters also results in a higher concentration of AGN feedback. Massive haloes lead to enhanced black hole accretion and undergo numerous mergers, which results in larger black holes. As pointed out by \citet{Henriques2017}, this produces a correlation between environment and AGN feedback.  AGN feedback occurs earlier and more frequently in protoclusters because of the higher abundance of massive haloes.

The protocluster halo mass function both enhances galaxy growth and suppresses star formation, and the competition between these processes result in the different SFH and stellar mass assembly of cluster galaxies compared to field galaxies.  Fig.\,\ref{fig:sm_total}c shows that the processes causing efficient star formation dominate at $z>1$, whilst quenching and disruption processes dominate at later times, resulting in inefficient star formation. 

\subsection{Testing galaxy formation with observations of protoclusters}

\subsubsection{The formation of stars}

The observed SFH of protoclusters presented in Fig.\,\ref{fig:CSF} qualitatively agrees with the simulation, but two results at odds with current observations and warrant further investigation. Fig.\,\ref{fig:CSF}b shows that the correlation between low SFRs and high density environments reduces with increasing redshift, but there is no reversal in the SFR-density relation. At $z>3$, the average sSFR of galaxies in both environments are comparable. However, several observational studies of high redshift clusters have claimed that there is a reversal, such that star formation is enhanced in clusters relative to the field \citep[e.g.][]{Tran2010,Santos2015}. Such a reversal would mean that a physical process is missing from our theory of galaxy formation, so it is vital that the SFRs and sSFRs of protoclusters are robustly determined. 

One of the biggest observational barriers to accurately measuring the SFR within high-redshift (proto)clusters is the difficulty of measuring dust-obscured star formation. Star forming galaxies are dustier in higher density environments \citep{Sobral2016b}, therefore a larger fraction of the star formation may be obscured in (proto)clusters than in the field.  The MIRI instrument on {\it JWST} and the Square Kilometer Array (SKA) will be able to resolve this issue as they will estimate SFRs using observables that are not attenuated by dust. 

\begin{figure}
 \psfrag{a}[][][1][0]{\Large${\rm log}[M_{200}/(h^{-1}{\rm M_{\odot}})]$}
 \psfrag{b}[][][1][0]{\Large${\rm log}[{\rm SFRD}/(h^{3}{\rm M_{\odot}yr^{-1}Mpc^{-3}})]$}
 \psfrag{c}[r][][1][0]{\Large$z=1.5$}
\includegraphics[width=90mm]{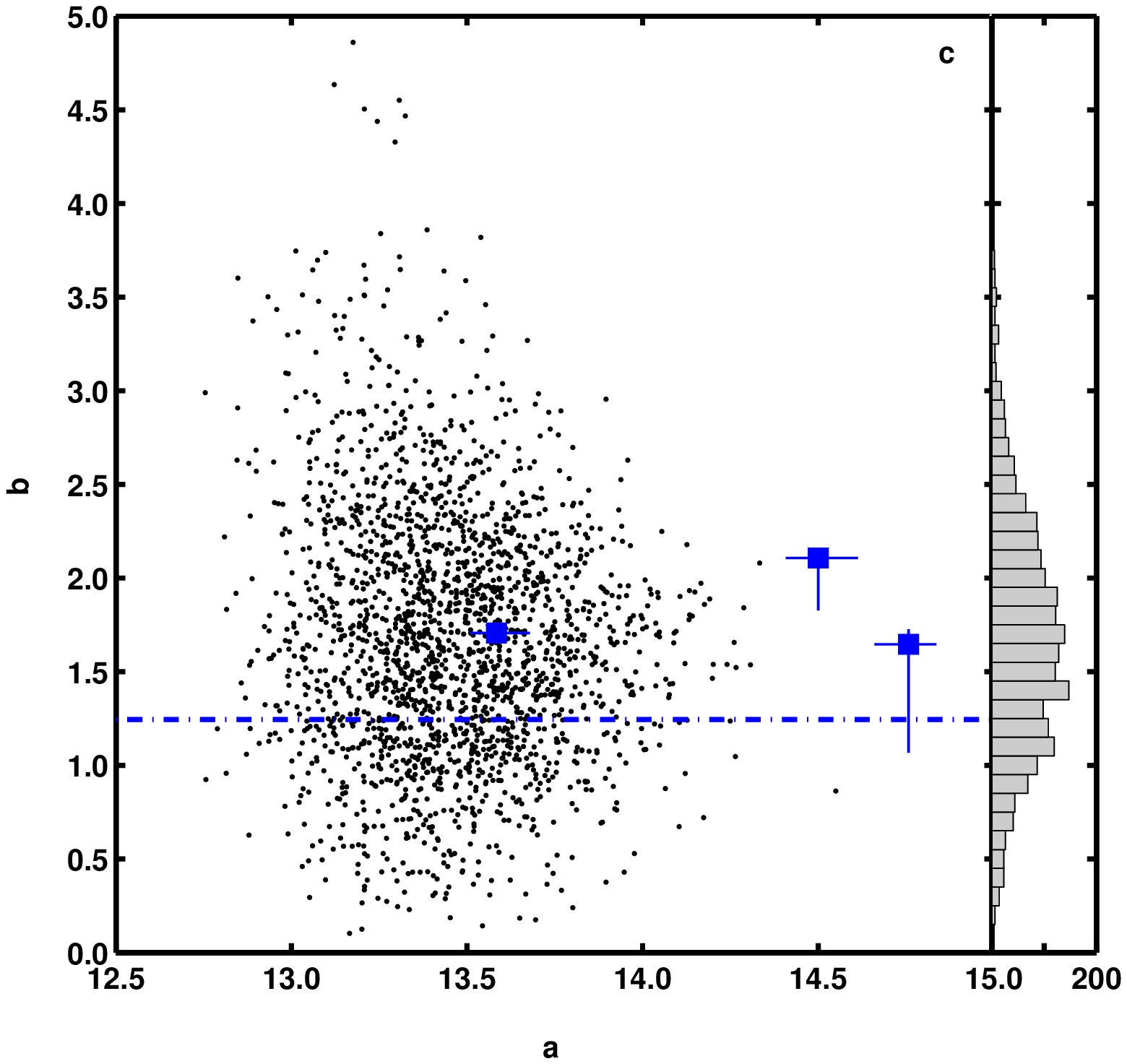}
\caption{\label{fig:SFRD}
The black circles show the simulated SFRD for all 2136 main haloes within the simulation box at $z=1.5$.  The large blue squares are observational data of clusters from \citet{Santos2013,Santos2014} and \citet{Santos2015} at $z=1.39$, 1.58 and 1.62. Although the three observed SFRDs of clusters are above the average SFRD of simulated main haloes (blue dot-dashed line), there are many main haloes that have similarly high SFRDs, therefore the observations are not in tension with the simulation.}
\end{figure}

The second apparent tension in Fig.\,\ref{fig:CSF} is the discrepancy between the observed and simulated SFRD in protoclusters and clusters. The model SFRDs in Fig.\,\ref{fig:CSF}c are calculated as the total SFR of all protocluster and main halo galaxies divided by the total volume of these regions within the simulation box, and is therefore an average of the SFRD in these environments. However, star formation may proceed in a haphazard and bursty manner which would not be reflected in the average evolution of the SFRD. In Fig.\,\ref{fig:SFRD} we show the individual SFRD for each simulated main halo at $z=1.5$ and overplot the observational data from \citet{Santos2013,Santos2014,Santos2015} at $z=1.39$, 1.58 and 1.62. Although the observational data points are above the average SFRD of the main haloes, they are well within the limits of individual simulated main haloes. 

The distribution of SFRD in Fig.\,\ref{fig:SFRD} shows that high-redshift main haloes of protoclusters will have SFRD ranging from a few to a few thousand ${\rm M_{\odot}yr^{-1}Mpc^{-3}}$, so passive and star-bursting clusters will co-exist at high redshift \citep[e.g.][]{Santos2015,Cooke2016}. This means that galaxy formation models cannot be tested with single protoclusters/clusters. Instead tens of protoclusters/clusters are required to sample the full distribution of star formation properties at each epoch. To measure the SFRD distribution we need a reasonably large sample of clusters and protoclusters which are not biased to rapidly star forming objects \citep[such as the {\it Herschel} selected sample of][]{Clements2014}. The ideal cluster sample would be selected from ICM detections, such as the samples from the {\it eRosita} satellite \citep{eRosita} and the South Pole Telescope (SPT), especially with the new SPT-3G detectors which will enable clusters to be detected beyond $z=1.5$ \citep{Benson2014}.

\subsubsection{Quenching of star formation}

The quenched fraction of field galaxies is used as a constraint for the \citet{Henriques2015} SAM so the simulation is designed to be in good agreement with the field observations. However, the QE and quenched fractions presented in Figs.\,\ref{fig:qe_cluster} and \ref{fig:Passfrac}, respectively, are in tension with recent observations of $z>1.6$ clusters.  \citet{Newman2014,Cooke2016} and \citet{Lee-Brown2017} show that the QE in high-redshift clusters is correlated with stellar mass at $M_{*}>10^{10}$\Msun. They find the most massive galaxies have a 100\% QE, whilst there is almost no environmental quenching acting on $M_{*}\sim10^{10.3}$\Msun\ galaxies. This is at odds with the simulation which shows an inverse correlation, or no correlation, with stellar mass at $z>1$. Furthermore, the observed quenched fractions of massive galaxies in these cluster studies are much higher than predicted by the simulation. On the other hand, the study of \citet{Nantais2017} reveals lower QEs that are more similar to the simulation, which suggests that there is a large amount of scatter in the QE between different clusters. 

Given the large scatter in observational results, samples of at least tens of distant clusters are required to measure the quenched fraction of cluster galaxies as a function of stellar mass. Further observations that confirm the discrepancy with the simulation would suggest that the environmental processes quenching protocluster galaxies are not sufficiently understood and modelled. Using environmentally dependent quenched fractions at $z>1$ to constrain the SAM may reveal more realistic quenching models. 

To accurately compare the observations to simulations we must ensure similar methods are used to separate star forming and quenched galaxies. Observations preferentially use $UVJ$ colour cuts \citep{Williams2009a} to make this selection, but the stellar population and dust models in simulations are not reliable enough to produce accurate galaxy colours \citep{Henriques2015}. While the stellar population models can be improved, it is the authors' opinion that observed passive fractions should be measured using sSFR measurements rather than $UVJ$ so they can be more reliably compared to simulations. 

\subsubsection{Galaxy assembly}
Protoclusters are excellent regions to observe hierarchical galaxy assembly because the importance of galaxy growth through mergers is increased relative to star formation in comparison to the field. This results in a different evolution of the protocluster GSMF compared to the field, as shown in Fig.\,\ref{fig:smf_num}. 
A quantitative measurement of the assembly of clusters galaxies can be gained from mapping the evolution of the protocluster GSMF and comparing it with simulations. Galaxy formation models will need to correctly determine merger rates as well as star formation histories in order to match the GSMF data in both the field and protoclusters, so using GSMFs from different environments as constraints for SAMs may allow us to break the degeneracy between galaxy growth by star formation and mergers.

There are several reasons why the evolution of the protocluster GSMF has not yet been measured. The sample size of well studied protoclusters is relatively small (a few handfuls at time of writing) hindering robust measurements of the normalisation and shape of the mass function. The future large dark energy surveys (e.g.\,LSST and {\it Euclid}) will solve this issue by detecting tens of thousands of clusters and protoclusters, but our results show that to robustly detect the environmental dependance of the GSMF requires complete mass functions to $M_{*}=10^{10}$\Msun, and preferably below. This is beyond the sensitivity limits of current protocluster surveys and the wide surveys of the forthcoming LSST and {\it Euclid} observatories, but should be possible with {\it JWST} NIRCAM observations and the ESO HAWK-I instrument \citep{Kissler-Patig2008} used in combination with the GRound Layer Adaptive optics Assisted by Lasers (GRAAL) facility \citep{Hibon2016}. 

The simulation shows that the evolution of the main halo GSMF provides information about the relative growth of BCGs (central) compared to halo growth (satellite population). The lack of evolution in the shape of the simulated GSMF of the main haloes suggests that the BCG growth rate is tightly correlated to the galaxy infall rate into the haloes at $z<2$, whereas the evolution at $z>2$ suggest the BCGs grow at a relatively faster rate due to rapid star formation. Measurements of the main halo GSMF are hindered by the difficulty of locating the main halo of protoclusters at high redshifts. {\it eRosita} and SPT-3G will identify high mass haloes/clusters up to $z<2$, but to identify evolutionary sequences we must identify lower mass haloes, and haloes at $z>2$. Our results suggest the main halo within protoclusters may be identified as hives of low mass, passively evolving galaxies. A key prediction of the SAM is that low mass galaxies  ($M_*<10^{9}\,h^{-1}$\Msun) are efficiently quenched in the main haloes. Fig.\,\ref{fig:Passfrac} shows that even at $z=4$, 40\% of $10^8\,h^{-1}$\Msun\ galaxies in the main haloes are classified as passive. This is quadruple the passive fraction in the field or the rest of the protocluster. This stark difference in the star forming properties of low mass galaxies may provide a method for identifying massive collapsed dark matter haloes at high redshift, and help isolate the location of the main halo within protoclusters. 

\section{Conclusions}
\label{sec:conclusions}
We have investigated the star formation and stellar mass assembly history of cluster galaxies using the \citet{Henriques2015} SAM. We have compared the model to observations of protoclusters and find qualitative agreement in the shape of the SFHs and stellar mass functions. We find that:

\noindent (i) Most of the stars and metals which end up in clusters formed within protocluster galaxies, well before the collapse of the cluster structure.   

\noindent (ii) The SFHs of protocluster and field galaxies differ: the SFR  peaks $\sim$0.7\,Gyr earlier and extends over a shorter period of time in protoclusters than in the field. This is due to enhanced quenching of star formation in protoclusters since at least $z=3$. 

\noindent (iii) Star formation is quenched in the massive haloes of protoclusters and primarily in the main halo. At $z>1$, low mass galaxies are quenched when they become satellites. Their hot haloes are tidally stripped and their cold gas reservoirs are either used up through forming stars, or removed by supernovae/stellar wind feedback. Without a hot halo they are unable to rejuvenate their gas supply and they become passive.  High mass galaxies at $z>1$ are quenched by AGN feedback. At $z<1$, galaxies are increasingly quenched by ram pressure stripping and gas exhaustion in addition to the aforementioned processes. 

\noindent (iv) Galaxies assembled differently in clusters compared to the field despite the mass functions being similar today.  Protocluster GSMFs were top-heavy relative to the field. The stellar matter within protoclusters formed earlier across a large number of small galaxies, and was gradually redistributed to larger galaxies and into the intra-cluster light.  This redistribution did not occur to such a great extent in the field where galaxies continued to assemble a higher fraction of their stellar mass through star formation.

The SFHs and stellar mass functions differ in protoclusters because dark matter haloes are biased tracers of the dark matter density field \citep{Tinker2010}. The simulation shows that dark matter haloes formed earlier in protocluster regions and the density of haloes increased as the protocluster collapsed. This resulted in a rapid rate of halo mergers, producing a top-heavy halo mass function. The simulation shows that the different halo mass distributions impact several baryonic processes, including the early and rapid formation of galaxies, the enhancement of AGN and subsequent quenching of star formation, increased galaxy merger rates, and increased disruption of low-mass satellites.

Protoclusters offer an alternative view of galaxy formation compared with surveys of the field, allowing us to investigate different physical processes that occur in these dense environments.  Our results show that the key observables that warrant further investigation are:
\begin{itemize}
\item the relative evolution of the sSFR of protocluster and field galaxies to determine whether star formation is enhanced in dense environments at $z>1$.
\item the distribution of SFRD of clusters and protoclusters at $z>1$ to determine the degree of stochastic star formation during cluster assembly.  
\item the quenching efficiency and quenched fraction of cluster/protocluster galaxies as a function of stellar mass to determine whether the environmental quenching processes employed in the galaxy formation models are realistic.
\item the evolution of the protocluster/main halo GSMF to determine the rate of galaxy growth through mergers.
\end{itemize}
\noindent  Galaxy formation models that can reproduce these environmentally-dependent star formation and mass constraints are likely to be more realistic. Therefore obtaining comprehensive observations of protocluster properties and adapting simulations to better match these observations would lead to a better understanding of galaxy formation and evolution.

\section*{Acknowledgements}
The authors wish to thank the anonymous referee for helping improve the quality of the manuscript.  SIM acknowledges the support of the STFC (ST/K001000/1).  NAH acknowledges support from STFC through an Ernest Rutherford Fellowship (ST/J002844/1). EAC acknowledges support from the ERC Advanced Investigator Grant DUSTYGAL(321334) and STFC (ST/L00075X/1 and ST/P000541/1). 
The Millennium Simulation databases used in this paper and the web application providing online access to them were constructed as part of the activities of the German Astrophysical Virtual Observatory (GAVO).
This research used the ALICE High Performance Computing Facility at the University of Leicester.




\bibliographystyle{mnras}\bibliography{References,mn-jour}

\bsp	
\label{lastpage}
\end{document}